%% file: sample-sigconf.tex
\documentclass[sigconf]{acmart}

\input{tex/macro}
\usepackage{multirow}
\usepackage{threeparttable}
\AtBeginDocument{%
  \providecommand\BibTeX{{%
    \normalfont B\kern-0.5em{\scshape i\kern-0.25em b}\kern-0.8em\TeX}}}

\copyrightyear{2022} 
\acmYear{2022} 
\setcopyright{acmcopyright}\acmConference[L@S '22]{Proceedings of the Ninth ACM Conference on Learning @ Scale}{June 1--3, 2022}{New York City, NY, USA}
\acmBooktitle{Proceedings of the Ninth ACM Conference on Learning @ Scale (L@S '22), June 1--3, 2022, New York City, NY, USA}
\acmPrice{15.00}
\acmDOI{10.1145/3491140.3528299}
\acmISBN{978-1-4503-9158-0/22/06}

\begin{document}
\fancyhead{}
\title[Laugh at Your Own Pace]{Laugh at Your Own Pace: Basic Performance Evaluation of \\Language Learning Assistance by Adjustment of Video Playback Speeds Based on Laughter Detection}

\author{Naoto Nishida}
\affiliation{%
  \institution{University of Tsukuba}
  \streetaddress{Tsukuba}
  \city{Ibaraki}
  \country{Japan}}
\email{nawta1998@gmail.com}

\author{Hinako Nozaki}
\affiliation{%
  \institution{University of Tsukuba}
  \streetaddress{Tsukuba}
  \city{Ibaraki}
  \country{Japan}}
\email{nozaki@iplab.cs.tsukuba.ac.jp}

\author{Buntarou Shizuki}
\affiliation{%
  \institution{University of Tsukuba}
  \streetaddress{Tsukuba}
  \city{Ibaraki}
  \country{Japan}}
\email{shizuki@cs.tsukuba.ac.jp}

\renewcommand{\shortauthors}{Nishida, et al.}

\begin{abstract}
  Among various methods to learn a second language (L2), such as listening and shadowing, \textit{Extensive Viewing} involves learning L2 by watching many videos. 
  However, it is difficult for many L2 learners to smoothly and effortlessly comprehend video contents made for native speakers at the original speed. 
  Therefore, we developed a language learning assistance system that automatically adjusts the playback speed according to the learner's comprehension.
  Our system judges that learners understand the contents if they laugh at the punchlines of comedy dramas,
  and vice versa.
  Experimental results show that this system supports learners with relatively low L2 ability (under 700 in TOEIC Score in the experimental condition) to understand video contents. 
  Our system can widen learners' possible options of native speakers' videos as \textit{Extensive Viewing} material.
\end{abstract}



\begin{teaserfigure}
  \centering
  \includegraphics[width=0.95\textwidth]{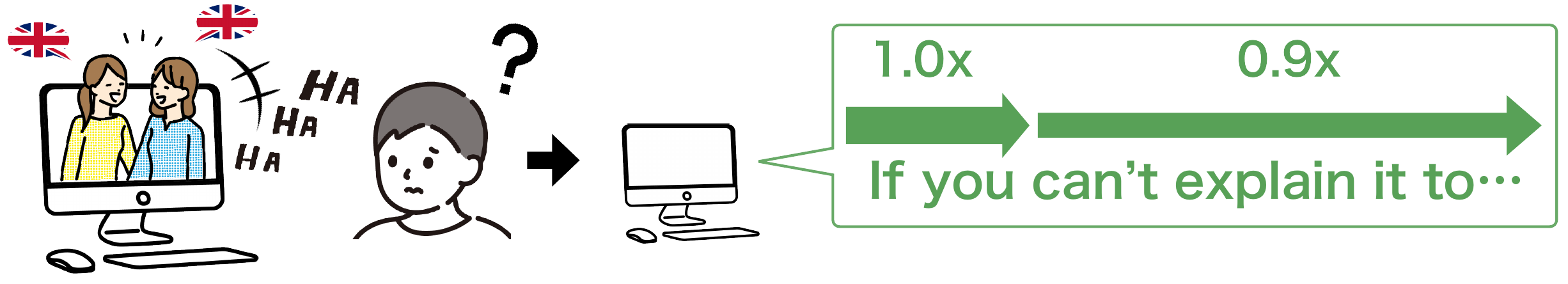}
  \caption{System overview. If learners do not laugh at the punchlines of videos, our system judges that they do not comprehend the video content, and gradually decreases the playback speed. Conversely, the speed increases if learners laugh.}
  \Description{System overview.}
  \label{fig:teaser}
\end{teaserfigure}

\maketitle

\section{Introduction}
\input{tex/01_introduction}

\section{Related Work}
\input{tex/02_relatedwork}


\section{Implementation}
\input{tex/04_implementation}

\section{Experiment}
\input{tex/05_Experiment}

\section{Discussion \& Future Work}
\input{tex/06_discussion_and_futurework}

\bibliographystyle{ACM-Reference-Format}
\bibliography{sample-sigconf}

\end{document}

%% file: tex/macro.tex
\newcommand{\figref}[1]{Figure~\ref{#1}}
\newcommand{\tbref}[1]{Table~\ref{#1}}

%% file: tex/01_introduction.tex
Among various methods to learn a second language (L2), 
Renandya and Jacobs defined \textit{Extensive Viewing} as ``listening to large amounts of motivating and engaging materials which are linguistically appropriate over a while where they listen with a reasonable speed for general understanding, with a focus on meaning rather than form''~\cite{willy2016}.
Ivone and Renandya further defined it as ``to be exposed to a large amount of easily comprehensible and enjoyable materials presented in the target language over an extended period'' as a kind of \textit{Extensive Listening}~\cite{ivone2019}.
\textit{Extensive Viewing} has various benefits as a learning method;
learners can improve their listening fluency~\cite{willy2016, chang2016},
increase their listening vocabulary~\cite{wang2012, masrai2020, feng_webb_2020},
and get used to liaisons, slang, and spoken forms of the target language~\cite{renandya2011, betul2014}.
Besides, it is also superior to akin \textit{Extensive Learning} methods because engaging and real-life videos can maintain learners' motivation towards learning~\cite{ozgen2020}, 
learners can start learning whenever they want~\cite{webb2015, hungchun2020}, 
and captions can aid the comprehension for the contents~\cite{peters2016, hayati2011, lee2013}. 
Therefore, 
\textit{Extensive Viewing} recently gained the attention of language learners as a valuable and enjoyable method~\cite{ivone2019}.

However, 
it is difficult for many L2 learners to view the learning materials at the rate at which native speakers do because of their undeveloped reading and listening knowledge of the target language~\cite{webb2015, Blau1990}.
As such, 
they usually face scenes where they cannot fully understand the content at the original playback speed.

As easily comprehensible material is essential for \textit{Extensive Viewing}~\cite{ivone2019}, 
video materials for native speakers (e.g., drama series) at the original speed are not suitable for L2 learners.

On the other hand,
making L2 learners manipulate interfaces to change the playback speed is also problematic. 
This is because continuous manipulation over a long period can make learners less motivated~\cite{fujii2019}, 
which is not suitable for \textit{Extensive Viewing} as a method that takes over months or years~\cite{willy2016,webb2015, mori2015}.

Therefore, 
we created a language learning assistance system that enables L2 learners to effortlessly engage in long-lasting \textit{Extensive Viewing} of videos made by native speakers.
This system can automatically adjust the playback speed of videos depending on the comprehension of each L2 learner based on their facial expressions (\figref{fig:teaser}).

For this purpose,
we first examined if learners act differently depending on their comprehension.
From our preliminary study, 
we found that if learners laugh at the punchlines of videos, then they comprehend the content, and vice versa.
We then proved that our proposed system is adequate for L2 learners with low proficiency.
Our research can broaden learners' possible choices of materials for \textit{Extensive Viewing}.

Our contributions can be summarized as follows:
\begin{itemize}
\vspace{-0mm}
    \item We found appropriate indicators that suggest learners' comprehension at a specific time while viewing videos (i.e., gaze, facial expressions, and other gestural behaviors).
    \item We developed a novel method to support L2 learners by adjusting the playback speed according to the comprehension of each learner.
    \item We evaluated the system via an experiment, revealing that it is beneficial for learners whose L2 skills are relatively low (under 700 in TOEIC score in the experimental condition).
\end{itemize}
\vspace{-3mm}




%% file: tex/02_relatedwork.tex
\subsection{Existing Computer-Assisted Language Learning Methods}
In the context of \textit{Extensive Viewing}, 
existing computer-assisted language learning methods mainly focus on developing the vocabulary sizes of L2 learners.
For instance, 
Hu et al. and Sakunkoo et al. developed a system that enables learners to watch videos with interactively translated captions,
where learners hover a mouse on unknown L2 words~\cite{sathaporn2018,sakunkoo2013}. 
Fujii and Rekimoto developed a system that detects English learners' proficiency level to suggest a moderate number of translations~\cite{fujii2019}.

However, 
there are few systems to support L2 learners using playback speed control in \textit{Extensive Viewing}.
Besides, 
existing systems require a series of movements via mouse cursors from users to change the speed of videos~\cite{sathaporn2018, languagereactor, subtitlesforll}.
\textit{Extensive Viewing} requires continuous efforts towards learning from learners.
Therefore, 
the workload necessary for viewing should be reduced as much as possible.

Thus, 
we developed a system that supports L2 learners through video speed control and requires less workload.

\subsection{Relationship between Playback Speed and Learner Comprehension}
In terms of foreign language learning, 
Blau and Sugai et al. confirmed that a slower speech rate facilitates listening comprehension among L2 learners~\cite{Blau1990, sugai2016}.
On the other hand, 
in terms of Massive Open Online Courses (MOOCs) in mother tongues, 
Kao et al. confirmed that sped up lecture videos boost MOOC learners' comprehension, 
while shortening the viewing time and raising the probability of completing the course~\cite{kao2014, lang2020}.

From these results, 
we induced that if learners cannot understand the video contents well, 
slowing the video content would be helpful because they can spend additional time processing the oral information, 
whereas 
if learners can understand them well, 
speeding up the video content would benefit learners because they can spare unnecessary time to watch and keep their concentration.

\subsection{Reduction of Learners' Workload}
For learners maintaining a daily habit of \textit{Extensive Viewing}, 
the workload necessary for the manipulation should be reduced as much as possible.
Fujii et al. and Song et al. used unconscious gaze movements or postures to predict learners' L2 proficiency levels~\cite{fujii2019, song2015}.
Similarly, 
Arakawa and Yakura used learners' unconscious reactions towards oral alterations to keep their attention during video-based learning, 
while maintaining the mental workload at a constant level~\cite{arakawa_yakura2021}. 

Hence, 
we used learners' unconscious behavior to alleviate the workload necessary for learning. 
Specifically, 
we focused on whether learners laugh when they face the punchline scenes of video material.

\subsection{Apparatus for \textit{Extensive Viewing}}
One of the aims of \textit{Extensive Viewing} is to increase the L2 learning time outside of the classroom~\cite{webb2015, willy2016}. 
In addition, 
\textit{Extensive Viewing} requires long-lasting continuation on a daily basis~\cite{webb2015}.
Therefore, 
to reduce the workload required for the setup for learning anywhere and anytime, 
we should minimize the apparatus for \textit{Extensive Viewing}.

Therefore, 
we examined features that we can capture via a built-in webcam of an ordinary laptop computer (MacBook Pro). 


%% file: tex/04_implementation.tex
\begin{figure}
    \centering
    \includegraphics[width=70mm]{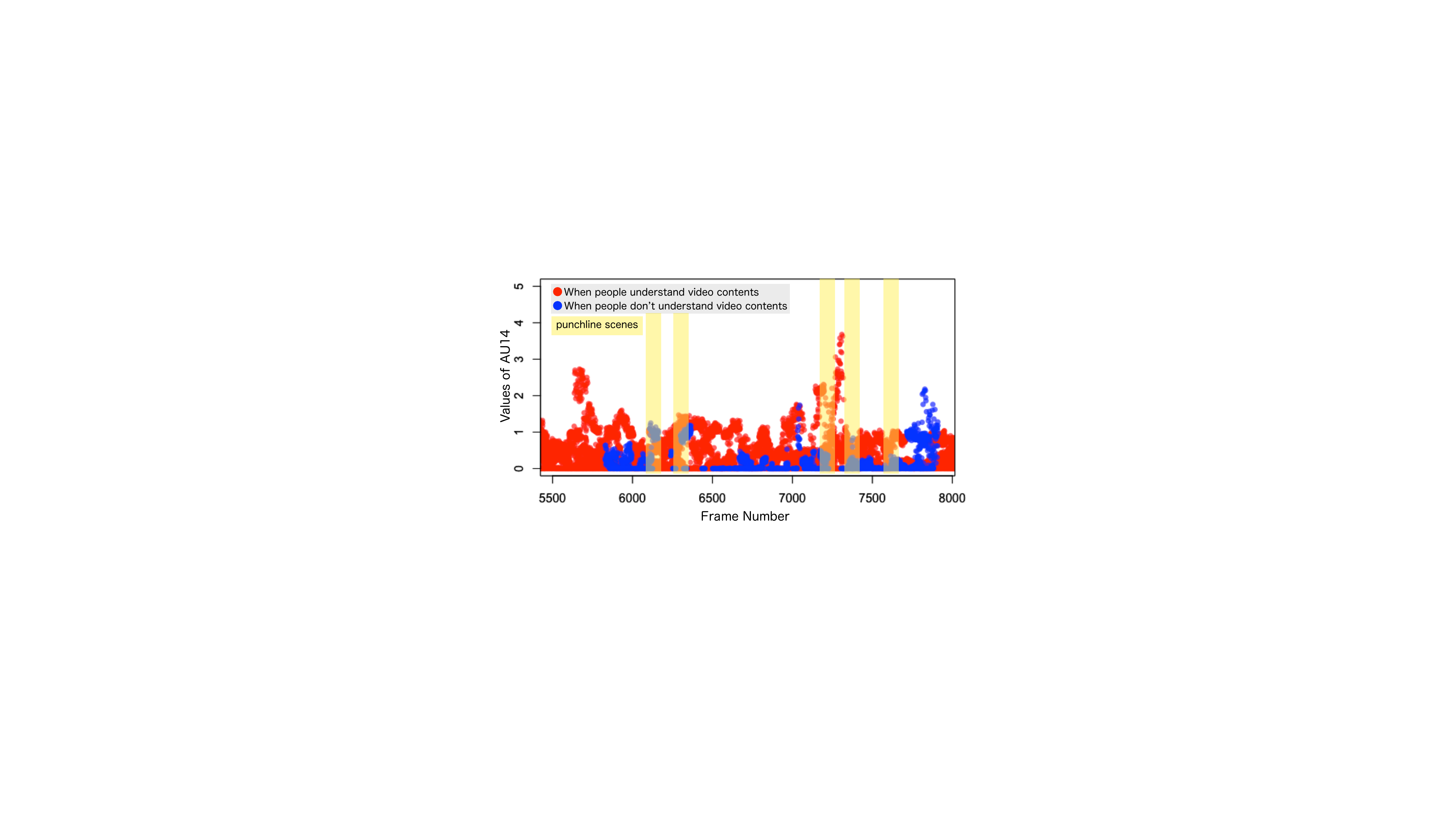}
    \caption{Values of Action Units 14 (\textit{AU14}) of the 5 participants according to frame number. Red points represents the case when learners understand the contents. Blue points represents the case when they do not understand. Yellow ranges are punchline scenes in the experimental video. In the preliminary study, participants watched the video twice. First, they watched it normally; second, while watching they classified when they can keep up with the story and when they cannot.}
    \label{fig:AU14}
\end{figure}

Through our preliminary study with 5 participants (all males, M=23.6, SD=1.82, all of them are Japanese and learn English as a foreign language), 
we discovered that people laugh at the punchlines of content only when they understand the English video content well.
\figref{fig:AU14} presents the total Action Unit 14 (AU14) values according to the frame number of the experimental videos\footnote{\url{https://drive.google.com/file/d/1dUrHqrXmw1APNtyVhdz-2tVeLOofoMbN/view?usp=sharing}}.
Action Units are measures to capture human facial movements by the appearance on their face~\cite{ekman1978, ekman1993, ekman2007}.
AU14 corresponds to a movement of buccinator, 
which can be an index of smile and laugh~\cite{mehdi2016, tian2001}.
In the analysis of our preliminary study, 
we confirmed a significant difference in AU14 between when learners keep up with the contents and when they cannot (p<.001 in Mann-Whitney U test, and effect size 0.22 in Hedges' g; we used Kolmogorov-Smirnov test for normality confirmation), 
whereas the other AUs, face tilt, and gaze direction did not show a significant difference.
Some participants also said that if they cannot keep up with the story, 
their face gets strained, and they cannot laugh at the punchlines of the contents.
Additionally, 
we prototyped a system that adjusts the playback speed of videos depending on the viewer's facial expressions and demonstrated it at a workshop in Japan~\cite{nishida2021en}.
We then ensured changing the playback speed could help learners comprehend the contents.

Based on the result of that preliminary study, 
we implemented a language learning assistance system that personalizes the playback speeds of video material based on laughter detection in the punchlines of the videos.

We used Python\,3.7.4, Perception for Autonomous Systems (PAZ) library for facial expression detection, Selenium library for manipulation of the video player, and Laughter-Detection library for detecting the punchlines of video materials~\cite{octavio2021, selenium2021, gillick2021}.
We defined a punchline as a scene where a laugh track is inserted, 
assuming that situation comedy (i.e., content with laugh tracks) is the de facto standard for \textit{Extensive Viewing}~\cite{ozgen2020, webb2015}.
We chose video streaming sites (e.g., YouTube, Netflix) as a video player platform because these sites employ a time stretch algorithm, which provides a steady pitch regardless of playback speed changes ~\cite{youtube_timestretch}.

This system decreases the playback speed by 0.1x when the user does not show a smile or laugh in punchline scenes in the videos, 
and increases the speed by 0.1x when the user shows a smile or laugh in punchline scenes.
The playback speed ranges from 0.6x to 1.0x (original speed). 

%% file: tex/05_experiment.tex
We conducted an experiment to evaluate the system's basic performance to help learners understand videos while keeping the workload and enjoyment towards the content steady, 
which are essential factors for the continuation of \textit{Extensive Learning}. 
We selected English as the target language because of its prevalence, 
making it easier to look for participants who learn English as a foreign language (EFL). 

\subsection{Participants}
We recruited EFL learners whose English proficiency level is above B1 in CEFR score~\cite{cefr2001}, 
which is 550 in TOEIC score~\cite{tannenbaum2008}.  
This is because \textit{Extensive Learning} is a method mainly for people who know at least 3000 basic words~\cite{webbrodgers2009b, pujadas2020, rodgers2013}, and people whose English CEFR score is above B1 generally know more than 3000 words~\cite{james2009, staehr2008, velonica2019}. 
Twenty participants (4 females, 16 males, M=22.6, SD=1.59, all Japanese) participated in this experiment.  
We asked the participants to join this experiment with their ordinary eye conditions and make their faces easy to see, such as brushing up bangs, during the experiment. 
8 participants joined the experiment with naked eyes, 
5 participants with glasses, 
and 7 participants with contact lenses. 

\subsection{Apparatus and Location}
We used a laptop PC (MacBook Pro 13 inch, Big Sur 11.6, 2.8\,GHz quad-core Intel Core i7, 16\,GB memory). 
The built-in webcam was a 720p FaceTime HD camera (1280 x 720\,pixels). 
We avoided backlit conditions to capture the face clearly. 
The distance between a participant's face and the PC display was from 40\,cm to 60\,cm, 
which is the recommended distance of PC manufacturer~\cite{distance}.

\begin{figure}
    \includegraphics[width=75mm]{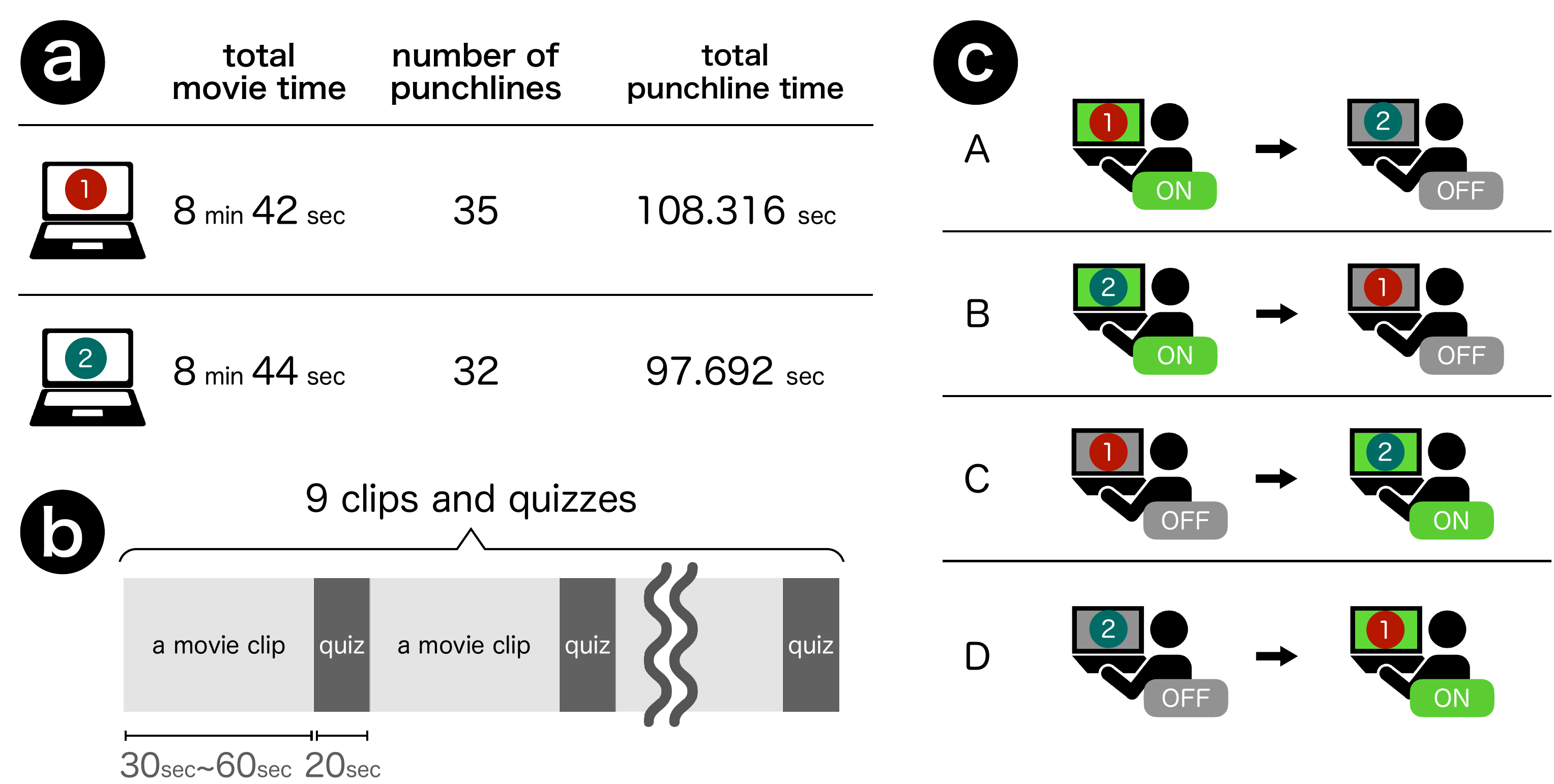}
    \caption{Control configuration: a) control in two videos, b) composition of a video, and c) control in groups.}
    \label{fig:control_config}
\end{figure}

\begin{table}[]
\begin{center}
\begin{threeparttable}
\caption{TOEIC scores in each group.}
\vspace{-3mm}
\label{tab:toeic_score}
\begin{tabular}{|l|lllll|l|l|}
  \hline
  & \multicolumn{5}{l|}{TOEIC score}   & Mean (SD)                   \\ \hline
A & 565  & 635  & 700 & 755 & 855 & 702.00 (111.2)    \\ \hline
B & 550  & 690  & 755 & 760 & 815 & 714.00 (101.8)    \\ \hline
C & 590  & 670  & 730 & 790 & 845 & 725.00 (99.87)    \\ \hline
D & 550\tnote{\dag} & 670  & 680 & 785 & 880 & 713.00\, (111.9) \\ \hline
\end{tabular}
\begin{tablenotes}
  \item[\dag] One person had not taken TOEIC. Thus, we hypothesized their score from their statement.
\end{tablenotes}
\end{threeparttable}
\end{center}
\vspace{-4mm}
\end{table}

\subsection{Configuration}
To reduce bias from the video itself, 
we made two similarly constructed videos (\textit{V1}\footnote{\url{https://drive.google.com/file/d/1LYx2rFDSLjlLNZo5Pa6qiUuhSweM_Y0W/view?usp=sharing}} and \textit{V2}\footnote{\url{https://drive.google.com/file/d/18cyV69l3x0ieA6xVUyg15MuHxcDOJKMz/view?usp=sharing}}).
578 words were spoken in \textit{V1}, 
and 469 words were spoken in \textit{V2}.
Each video was composed of 9 clips from a sitcom \textit{Friends} and 9 corresponding comprehension quizzes about the clips (\figref{fig:control_config}\,a and \figref{fig:control_config}\,b).
Each clip's length was approximately from 60 to 90 seconds, and each quiz's length was 20 seconds based on the example of listening quizzes of TOEIC and TOEFL~\cite{toefl_official,toeic_official}.
Because we compare the difference in comprehension between when a participant uses our system and when they do not, 
we divided the participants into 4 groups (\figref{fig:control_config}~c,  \tbref{tab:toeic_score}).
We allocated them according to their TOEIC score because the reliability of the test is high (0.90)~\cite{im2019test, kuder1937theory}, 
meaning that its score almost certainly reflects learners' L2 ability.
In addition, 
to reduce the effects of the participants' vocabulary ability, we aimed to ensure every word in the video was known.
To this end, 
we let them read a series of quizzes shown in the videos and taught them the unknown vocabulary spoken in the videos.
We also had them read a complex vocabulary list. 
We took the 20 most rarely seen words from each video with reference to the \textit{Corpus of Contemporary American English}~\cite{coca}. 
We ensured all participants knew the vocabulary in the videos and quizzes in our post-questionnaire.

\vspace{-1mm}

\subsection{Procedure}
We take group A (\figref{fig:control_config}~c) as an example to explain the procedure. 
First, 
participants answered a vocabulary test (\textit{VT1}) whose vocabulary appears in the following \textit{V1}. 
The vocabulary test was conducted to teach participants the difficult words necessary to keep up with the story; 
therefore, 
we told participants the correct answer immediately after they finished answering. 
Second,
we let them read a series of quizzes (\textit{Q1}) to be presented in the quiz time in \textit{V1} to ensure they had no vocabulary problems while answering. 
Third, 
participants viewed \textit{V1} with our proposal system, answering \textit{Q1} at the same time. 
Fourth, 
participants answered \textit{VT1} again to be checked if they had no vocabulary issues while viewing \textit{V1}.
Fifth, 
participants answered NASA-TLX, an assessment tool that rates perceived workload~\cite{hancock1988}.
After that, 
participants were given the same series of workflows again under the condition of \textit{V2}, \textit{VT2}, \textit{Q2}, and without-system viewing instead of \textit{V1}, \textit{VT1}, \textit{Q1}, and with-system viewing.
Participants were given 5 minutes rest after answering NASA-TLX.
At the end of the experiment, they answered a free answer sheet and SUS to assess usability~\cite{brooke1996}. We told the participants to write about any unknown words in the videos and quizzes.
The whole experiment time was 64 minutes on average.
Whether a participant used the system was single-blinded. 

\begin{table}[]
  \caption{Time spent on each process.}
  \label{tab:answer_time_in_quiz_vocabtest}
  \footnotesize
\begin{tabular}{|l|lllllll|}
\hline
\multirow{3}{*}{} & \multicolumn{4}{l|}{answering time on \textit{VT1}, \textit{VT2}}                                                           & \multicolumn{2}{l|}{\multirow{2}{*}{\begin{tabular}[c]{@{}l@{}}reading quiz \\ beforehand\end{tabular}}} & \multirow{3}{*}{\begin{tabular}[c]{@{}l@{}}viewing time\\ with the system\end{tabular}} \\ \cline{2-5}
                  & \multicolumn{2}{l|}{pre-viewing}                    & \multicolumn{2}{l|}{post-viewing}                   & \multicolumn{2}{l|}{}                                                                                    &                                                                                         \\ \cline{2-7}
                  & \multicolumn{1}{l|}{\textit{VT1}} & \multicolumn{1}{l|}{\textit{VT2}} & \multicolumn{1}{l|}{\textit{VT1}} & \multicolumn{1}{l|}{\textit{VT2}} & \multicolumn{1}{l|}{\textit{Q1}}                             & \multicolumn{1}{l|}{\textit{Q2}}                            &                                                                                         \\ \hline
mean(s)         & 111.8                   & 138.6                   & 44.86                    & 48.37                    & 169.6                                              & 164.0                                             & 634.5                                                                                  \\ \hline
SD(s)           & 31.27                    & 56.57                    & 13.02                    & 19.51                    & 86.79                                               & 92.38                                              & 44.62                                                                                   \\ \hline
\end{tabular}
\vspace{-3mm}
\end{table}



\begin{figure}
    \includegraphics[width=80mm]{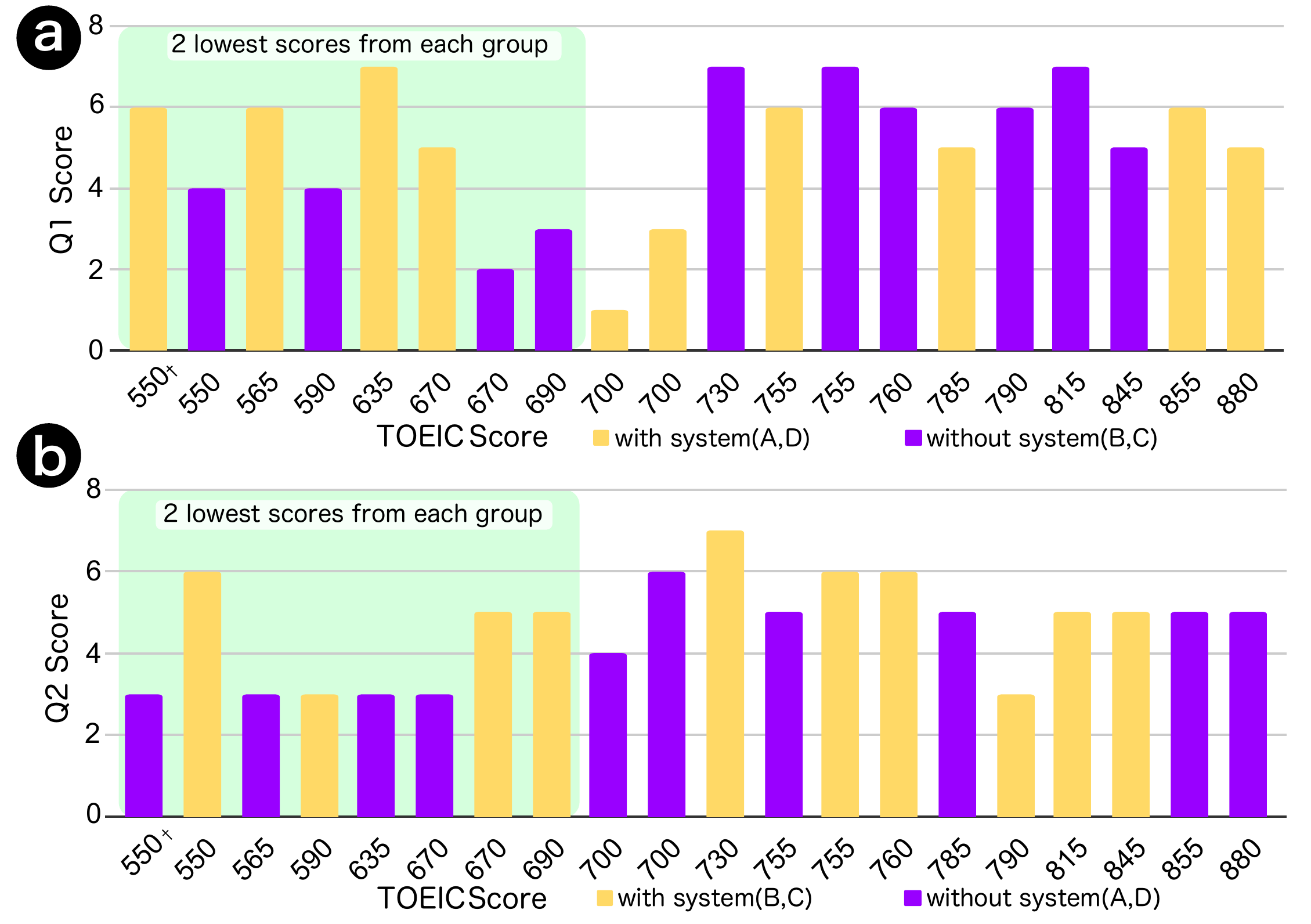}
    \caption{Relationship between TOEIC score and \textit{Q1} (a) or \textit{Q2} (b). Relatively, participants with low TOEIC score had a higher score when they used our system. The green area indicates the participants with the lowest and second lowest scores from each group.}
    \vspace{-3mm}
    \label{fig:comprehension_quiz_compiled}
\end{figure}
\vspace{-4mm}

\begin{figure}
    \includegraphics[width=85mm]{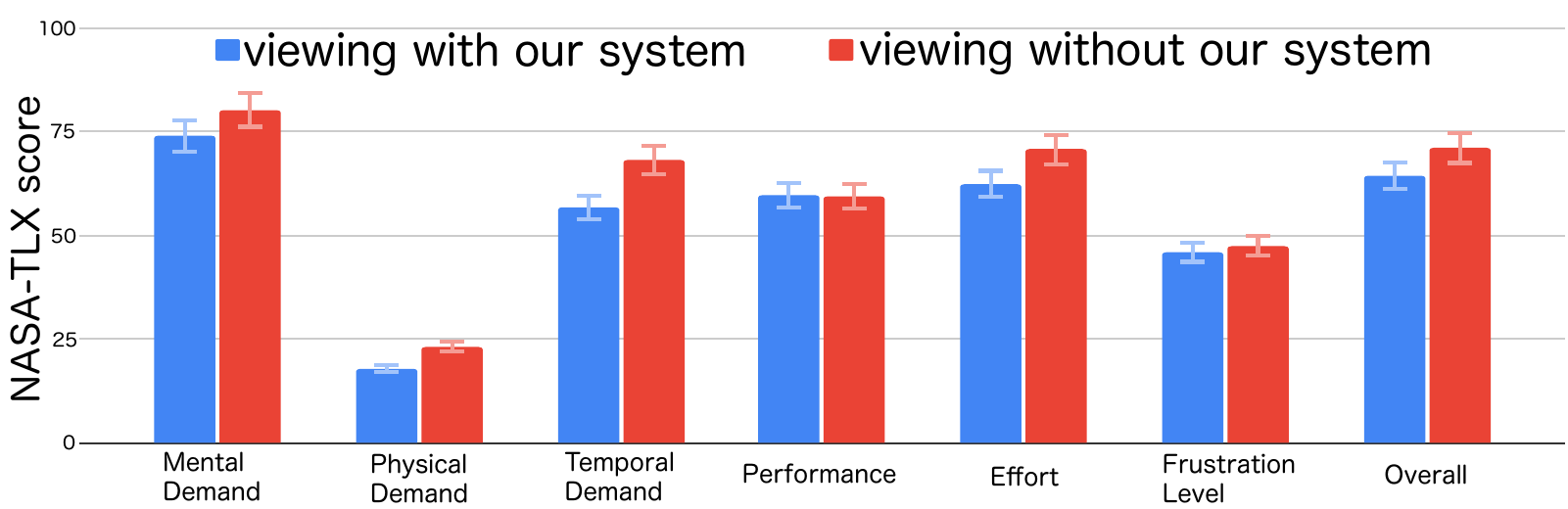}
    \caption{NASA-TLX score and its 10\% error range. This shows that our system mostly does not worsen learners' mental workload. Besides, it significantly alleviates \textit{Temporal Demand} and \textit{Effort}.}
    \vspace{-3mm}
    \label{fig:nasatlx}
\end{figure}

\subsection{Results}
The time spent on each process is shown in \tbref{tab:answer_time_in_quiz_vocabtest}.
The time participants viewed videos without the system was 523 seconds on average, meaning they spent 1.2 times longer than they viewed without the system.

Regarding the participants' comprehension,
as shown in \figref{fig:comprehension_quiz_compiled},
learners with low L2 proficiency (the lowest 2 people from each group) could answer both \textit{V1} and \textit{V2} more correctly with our system (\textit{Q1}:p<.005 and \textit{Q2}:p<.05, the effect sizes are 2.69 and 1.71 respectively). 
This means that learners with low L2 proficiency levels can understand videos more with our system. 

In terms of mental workload,
NASA-TLX shows lower scores from perspectives other than \textit{Performance}~\cite{hancock1988}, showing significant differences in \textit{Temporal Demand} and \textit{Effort}~(\figref{fig:nasatlx}).  

As for system usability,
SUS scored 73.75 on average (SD=14.66), which represents \textit{Good} usability~\cite{brooke1996}. 

%% file: tex/06_discussion_and_futurework.tex
From our experiment, 
we confirmed that our system enhanced L2 learners' comprehension of video content as a language learning material. 
The effect was particularly significant for learners with low proficiency (under 700 in TOEIC score in our experiment condition). 
We also proved that this system costs learners less mental workload and provides \textit{Good} usability.

In future work,
we will increase the number of participants with lower L2 proficiency (B1 or below in L2 proficiency).
We also plan to discover more appropriate ways to extract learners' comprehensions.
Besides, 
we will improve our system so that people can speed up the playback rate past 1.0x to skip any scene they can easily understand. 
This allows learners to spare time, leading to less workload for learning and the daily continuation of \textit{Extensive Viewing}.